\documentclass[aps,prd,twocolumn,showpacs,groupedaddress,nofootinbib]{revtex4}
\usepackage{graphicx}

\begin{document}

\title{Estimate of the charmed $0^{--}$ hybrid meson spectrum from quenched lattice QCD}

\author{Yan Liu}
\address{Department of Physics, Zhongshan (Sun Yat-Sen) University,
Guangzhou 510275, China}

\author{Xiang-Qian Luo}
\thanks{Corresponding author. Email address: stslxq@mail.sysu.edu.cn}
\address{
CCAST (World Laboratory), P.O. Box 8730,
Beijing 100080, China\\
Department of Physics, Zhongshan (Sun Yat-Sen) University,
Guangzhou 510275, China}
\thanks{Mailing address.}

\date{\today}

\begin{abstract}
We compute from quenched lattice QCD the ground state masses of
the charmed hybrid mesons ${\bar c} c g$, with exotic quantum
numbers $J^{PC}=1^{-+}$, $0^{+-}$ and $0^{--}$. The $0^{--}$
hybrid meson spectrum has never been provided by lattice
simulations due to the difficulties to extract high gluonic
excitations from noise. We employ improved gauge and fermion
actions on the anisotropic lattice, which reduce greatly the
lattice artifacts, and lead to very good signals. The data are
extrapolated to the continuum limit, with finite size effects
under well control. For $1^{-+}$ and $0^{+-}$ hybrid mesons, the
ground state masses are 4.405(38) GeV and 4.714(52) GeV. We
predict for the first time from lattice QCD, the ground state mass
of $0^{--}$ to be 5.883(146) GeV.
\end{abstract}


\maketitle

A hybrid (exotic) meson ${\bar q} qg$ is a bound state of quark
$q$, anti-quark ${\bar q}$ and excited gluon $g$. The excited
gluon makes quantum number of the bound state to be $1^{-+}$,
$0^{+-}$ or $0^{--}$, ......, inaccessible to $\bar{q}q$ mesons in
the quark model. The existence of hybrid mesons is one of the most
important predictions of quantum chromodynamics (QCD).

So far no signal for heavy exotic hybrid mesons has been
experimentally observed, though a number of potential candidates
for light hybrid mesons were
suggested\cite{Alde:1988qs,Adams:1998up}. Fortunately, this
situation may change, due to rapid development of new experiments,
for example PEP-$\amalg$(Babar), KEKB(Belle), 12 GeV Jefferson
Lab\cite{Szczepaniak:2003nu,Meyer:2003zq}, upgraded CLEO-c
detector\cite{Stock:2002uy}, and new BES3
detector\cite{Jin:2004ck}. Especially, 12 GeV Jefferson Lab,
CLEO-c and BES3 will present well needed and more reliable data
for the charmonium spectrum, including hybrid mesons.

The most reliable technique for computing hadron spectroscopy is
lattice gauge theory. It is a non-perturbative approach based on
first principle QCD. Of course, the lattice approach is not free
of systematic errors. The discretization errors in the Wilson
gluon and quark actions are the most serious ones. These errors
are smaller only at very small bare coupling, and very large
lattice volume is required to get rid of finite size effects. The
idea of Symanzik improvement\cite{Symanzik:1983dc} is to add new
terms to the Wilson actions to reduce the lattice spacing errors.
In combination with tadpole improvement\cite{Lepage:1992xa}, the
Symanzik program has recently led to great success in approaching
the continuum physics on very coarse and small lattices.
Simulations on anisotropic lattices help getting very good signal
in spectrum computations.

There have been many quenched lattice
calculations\cite{Griffiths:1983ah,Perantonis:1990dy,Bernard:1997ib,Lacock:1996ny,McNeile:1998cp,Manke:1998qc,Juge:1999ie,Drummond:1999db,Mei:2002ip,Bernard:2003jd,Hedditch:2005zf}
of the $1^{-+}$ or $0^{+-}$ hybrid meson masses, in either light
quark or heavy quark sector. The mass estimates for the light
hybrid mesons might still have some uncertainties, because those
simulations are still far from the chiral regime. The inclusion of
dynamical quarks is still very preliminary\cite{Bernard:2003jd},
due to limited computing resources. A recent review can be found
in Ref.\cite{Michael:2003xg}. It has been a long standing puzzle
for the $0^{--}$ hybrid mesons\cite{Bernard:1997ib}: no clear
signal has ever been found, which might be due to the fact that
the gluon is highly excited.

As for heavy quarks, special considerations have to be taken.
Currently, non-relativistic lattice QCD (NRQCD), and relativistic
heavy quark (Fermilab), and anisotropic relativistic approaches
are the leading methods. Let $a$ denote  the lattice spacing. The
NRQCD method\cite{Thacker:1991ej,Lepage:1992ej} is applicable for
$am_{q}>1$ and works well for very heavy quarks, especially for
the spin-independent $\bar{b}b$ system; however the continuum
limit is problematic because of the condition $am_{q}>1$; it is
difficult to include relativistic corrections and radiative
corrections, leading to breaking down of this method  for the
$\bar{c}c$ system\cite{Trottier:1997ej}. The relativistic
(Fermilab) approach to quarks\cite{El-Khadra:1997ej} works for
both light quarks and heavy quarks; Up to $O(a^{2})$, the
fermionic action is equivalent to the standard
Sheikholeslami-Wohlert(SW) action\cite{Sheikholeslami:1985ij}  on
an isotropic lattice; however, to get rid of the $O(a)$ error all
coefficients in the fermionic action are required to be
mass-dependent. The anisotropic relativistic approach to
quarks\cite{Klassen:1998fh,Okamoto:2001jb}, which is used in this
paper, generalizes the Fermilab approach to anisotropic lattice.
This improved quark action has been successfully applied to the
computation of the charmonium
spectrum\cite{Okamoto:2001jb,Chen:2000ej}, which agree very well
with experiments.

To investigate gluonic excitations in hadrons, additional
improvement of the gluon action would certainly help getting
better signals. The first attempt was made in Ref.
\cite{Mei:2002ip}, where the ground state masses of $1^{-+}$
hybrid mesons in the light quark and charm quark sectors were
computed, by combining the improved gluon
action\cite{Morningstar:1997prd} and a simplified relativistic
fermionic action\cite{Mei:2002ip} on the anisotropic lattice.
However, the quark masses were far away either from the chiral
limit or from the charm quark regime. The statistics were low, and
finite size effects and lattice spacing errors were not analyzed.

In this letter, we estimate the ground state masses of $1^{-+}$,
$0^{+-}$, and $0^{--}$ exotic mesons in the charm quark sector,
employing lattice QCD with tadpole improved
gluon\cite{Morningstar:1997prd} and
quark\cite{Klassen:1998fh,Okamoto:2001jb} actions on the
anisotropic lattice.  We get significantly improved signals for
these particles, in particular for the $0^{--}$ particle for the
first time.

\begin{table*}
\begin{center}
\tabcolsep 0.1in
\begin{tabular}{cccccccccccc}\hline
 $\beta$ & $\xi=a_s/a_t$ & $L^{3}\times T$   & $u_{s}$  & $u_t$  & $a_{t}m_{q0}$                   & $c_{s}$ & $c_{t}$ & $a_{s}$($1^{1}P_{1}-1S$)[fm]  & $La_{s}$[fm] & \\\hline
    2.6  &    3          & $16^{3}\times 48$ & 0.81921  &  1     & 0.229 $\tabcolsep 0.2in$ 0.260  & 1.8189  & 2.4414  & 0.1885(82)                    &  3.016       & \\
    2.8  &    3          & $16^{3}\times 48$ & 0.83099  &  1     & 0.150 $\tabcolsep 0.2in$ 0.220  & 1.7427  & 2.4068  & 0.1584(103)                   &  2.534       & \\
    3.0  &    3          & $20^{3}\times 60$ & 0.84098  &  1     & 0.020 $\tabcolsep 0.2in$ 0.100  & 1.6813  & 2.3782  & 0.1147(98)                    &  2.294       & \\
\hline
\end{tabular}
\end{center}
\caption{Simulation parameters at largest volume. We employed the
method in Ref.\cite{Okamoto:2001jb} to tune these parameters,
$\kappa_{t}$ and $\kappa_{s}$ for the quark action. The last two
columns are about the spatial lattice spacing and the lattice
extent in physical units, determined from the $1P-1S$ charmonium
splitting.} \label{tab1}
\end{table*}

Our simulation parameters are listed in Tab. \ref{tab1}. At each
$\beta=6/g^2$, three hundred independent configurations were
generated with the improved gluonic
action\cite{Morningstar:1997prd}. Two hundred configurations are
the minimum for obtaining stable results. We input two values of
bare quark mass $m_{q0}$ and then compute quark propagators using
the improved quark action\cite{Okamoto:2001jb}, and the hybrid
meson correlation function using the operators $1^{-+}=\rho\otimes
B$, $0_{P}^{+-}=a_{1}(P)\otimes B$, $0_{P}^{--}=a_{1}(P)\otimes E$
and $O_{S}^{--}=a_1\otimes E$ in Ref. \cite{Bernard:1997ib}, as
they give the best signals.

\begin{figure} [th]
\begin{center}
\includegraphics[totalheight=2.5in]{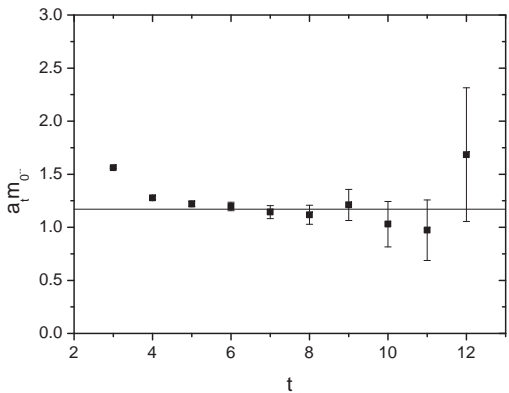}
\end{center}
\vspace{0.0cm} \caption{Effective mass of the $0^{--}$ hybrid
meson for $\beta=3.0$ and $a_{t}m_{q0}=0.100$. The solid line is
the fitted result, ranging from $t_i=6$ to $t_f=12$ with
$\chi^2/d.o.f.=0.4326$ and confidence level=0.7620.} \label{fig1}
\end{figure}

\begin{figure} [th]
\begin{center}
\includegraphics[totalheight=2.5in]{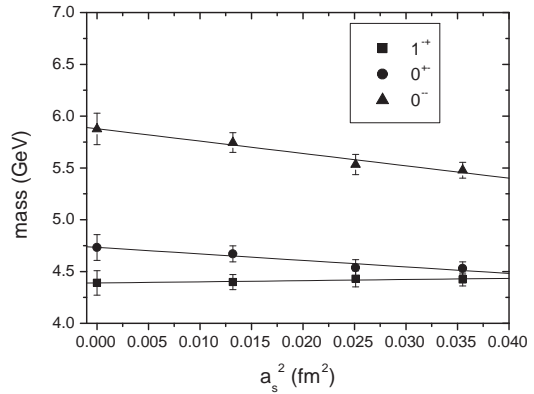}
\end{center}
\caption{Extrapolation of the charmed $1^{-+}$, $0^{+-}$, and
$0^{--}$ hybrid meson masses to the continue limit.} \label{fig2}
\end{figure}

\begin{figure} [th]
\begin{center}
\includegraphics[totalheight=2.5in]{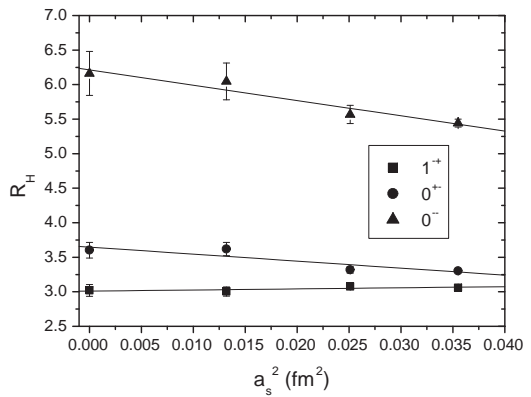}
\end{center}
\caption{Ratio of splittings $R_H=\Delta M(1H-1S)/\Delta
M(1^1P_1-1S)$ against $a_s^2$. The straight line is the
extrapolation to continuum limit.} \label{fig3}
\end{figure}

\begin{table*}
\begin{center}
\vspace{1.0cm} \tabcolsep 0.2in
\begin{tabular}{cccccc}\hline
 $\beta$       & $a_{s}^{2}(fm^{2})$ & $1^{-+}$     & $0^{+-}$    &$0^{--}$ & ~
 \\\hline
    2.6        & 0.0355             & 4.423(62)     & 4.530(63)   & 5.478(76) &~      \\
    2.8        & 0.0251             & 4.429(78)     & 4.536(79)   & 5.533(97)  &~     \\
    3.0        & 0.0132             & 4.398(73)     & 4.670(77)   & 5.745(95)  &~     \\
$\infty$       & 0                  & 4.390(118)    & 4.732(124)  & 5.876(152)      &~  \\
$\infty$       & 0                  & 4.405(38)     & 4.714(52)   & 5.883(146)   & $\leftarrow$ from  $R_H$ \\
 \hline
\end{tabular}
\end{center}
\caption{Charmed hybrid meson spectrum for the ground state. The
results in the continuum limit ($\beta=\infty$) were obtained by:
(i) directly extrapolating the data to $a_s^2 \to 0$; and (ii)
using the ratio of splittings $R_H$, as described in the text.}
\label{tab2}
\end{table*}

Figure \ref{fig1} shows an example of the effective mass plot
$\ln(C(t)/(C(t+1))$ of the $0^{--}$ hybrid, where $C(t)$ is the
correlation function between the hybrid operators. We obtained the
effective mass $a_tm$, with the fit range chosen according to
optimal confidence level and reasonable $\chi^2/d.o.f$.

We then interpolated the data to the charm quark regime using
$(m_{\pi}(\kappa_{t}\to\kappa_{t}^{charm})+3m_{\rho}(\kappa_{t}\to\kappa_{t}^{charm}))/4
\to M(1S)_{exp}=(m(\eta_{c})_{exp}+3m(J/\psi)_{exp})/4 =
3067.6$MeV, where the right hand side is the experimental value
for the $1S$ charmonium. The results for the charmed hybrid meson
masses are listed in Tabs. \ref{tab2} for the ground state. It is
also important to check whether these lattice volumes are large
enough. We also did simulations on $8^3\times48$ and $12^3\times
48$ at $\beta=2.6$, $12^3\times36$ at $\beta=2.8$, and $16^3\times
48$ at $\beta=3.0$, but here we just list the results from the
largest volume. When the spatial extent is greater than 2.2fm, the
finite volume effect on the $0^{--}$ mass is less than $0.1\%$ for
the ground state.

The $1^1P_1-1S$ charmonium splitting was chosen to determine the
lattice spacing, because it is roughly independent of quark mass
for charm and bottom sectors, and the experimental value $\Delta
M(1^1P_1-1S)_{exp}=457$MeV was well measured. Here we used
$\chi_{c_{1}}$ meson $(1^{++})$ mass for the P-wave and
$m(\eta_{c})/4+3m(J/\psi)/4$ for the S-wave. The results for the
spacial lattice spacing $a_s$ at different $\beta$ are listed in
Tab. \ref{tab1}. They are consistent with those from the heavy
quark potential\cite{Morningstar:1997prd}.

Figure \ref{fig2} shows the charmed $1^{-+}$, $0^{+-}$, $0^{--}$
hybrid meson masses as a function of $a^2_s$. Other hybrids have a
similar behavior, indicating the linear dependence of the mass on
$a^2_s$. The spectrum in the continuum limit is obtained by
linearly extrapolating the data to $a_s^2\to 0$, as also listed in
Tabs. \ref{tab2}.

We also computed the splitting of $1H-1S$ and the ratio
$R_{H}=\Delta M(1H-1S)/\Delta M(1^1P_1-1S)$, where $1H$ stands for
the ground state of a charmed hybrid meson. Its dependence on
$a_s^2$ and extrapolation to the continuum limit are shown in Fig.
\ref{fig3}. The last line of Tab. \ref{tab2} also lists the hybrid
meson masses for the ground state, using the following equation:
\begin{eqnarray}
\lim_{a_s^2 \to 0} M(1H)=&&M(1S)_{exp} \nonumber\\
&+&\Delta M(1^1P_1-1S)_{exp} \times \lim_{a_s^2 \to 0} R_H .
\nonumber
\end{eqnarray}
This method was claimed to be better\cite{Manke:1998qc}, because
the splitting between a hybrid and the $1S$ state is rather
insensitive to the imperfect tuning of $\Delta M(1^1P_1-1S)$ and
$M(1S)$. However, as seen in Tab. \ref{tab2}, the results from two
different methods agree very well.

One source of systematic errors in our calculation is the quenched
approximation. Although full QCD simulations will remove this
unknown error,  quenched approximation in some
areas\cite{Blum:2002ii,Gockeler:2003jf}, including the
hybrids\cite{Takahashi:2002it}, continues to play an important
role. The findings in Refs.
\cite{Michael:2003xg,Lacock:1998be,Manke:2001ft} indicate that the
effects of dynamical quarks on light hybrids and bottomed hybrids
are very small. To have full relevance of the charmed hybrids to
experiment, simulations with dynamical quarks, although extremely
expensive to achieve high statistics, might be helpful to see
whether the quenching error is under control. Nevertheless, our
results are a very important step for comparison with future
dynamical simulations.

Simulations on the anisotropic lattice with both gluon action and
improved quark action improved lead to the first observation of
the clear signal for the $0^{--}$ hybrids. We believe that our
findings are useful to experimental search for these new
particles, predicted by QCD.

\acknowledgments

We thank C. DeTar and E.B. Gregory for useful discussions. This
work is supported by the Key Project of National Science
Foundation (10235040), Project of the Chinese Academy of Sciences
(KJCX2-SW-N10) and Key Project of National Ministry of Eduction
(105135) and Guangdong Natural Science Foundation (05101821). We
modified the MILC code\cite{Milc} for simulations on the
anisotropic lattice. It has taken more than one year for the above
simulations on our AMD-Opteron cluster and Beijing LSSC2 cluster.


\begin{thebibliography}{99}

\bibitem{Alde:1988qs}
D.~Alde $\emph{et al.}$, Phys.\ Lett.\ B {\bf 205}, 397 (1988);
D.~Thompson $\emph{et al.}$, Phys.\ Rev.\ Lett.\ {\bf 79}, 1630
(1977); S.~Chung $\emph{et al.}$, Phys.\ Rev.\ D {\bf 60}, 092001
(1998); A.~Abele $\emph{et al.}$, Phys.\ Lett.\ B {\bf 423}, 175
(1998); $\emph{ibid.}$ \ {\bf 446}, 349 (1999).


\bibitem{Adams:1998up}
G.~S.~Adams $\emph{et al.}$, Phys.\ Rev.\ Lett.\ {\bf 81}, 5760
(1998).

\bibitem{Szczepaniak:2003nu}
  A.~Szczepaniak,
  %``Confinement and exotic meson spectroscopy at 12-GeV JLAB,''
  Braz.\ J.\ Phys.\  {\bf 33}, 174 (2003).
  %%CITATION = BJPHE,33,174;%%



\bibitem{Meyer:2003zq}
  C.~A.~Meyer,
  %``An experimental overview of gluonic mesons,''
  AIP Conf.\ Proc.\  {\bf 698}, 554 (2004).
%  [arXiv:hep-ex/0308010].
  %%CITATION = HEP-EX 0308010;%%


\bibitem{Stock:2002uy}
H.~Stock, arXiv:hep-ex/0204015.

\bibitem{Jin:2004ck}
  S.~Jin,
  %``Experimental review of new results on hadron spectroscopy,''
  Int.\ J.\ Mod.\ Phys.\ A {\bf 20}, 5145 (2005).
  %%CITATION = IMPAE,A20,5145;%%





\bibitem{Symanzik:1983dc}
K.~Symanzik,
%``Continuum Limit And Improved Action In Lattice Theories. 1. Principles And Phi**4 Theory,''
Nucl.\ Phys.\ B {\bf 226},  187 (1983);
%%CITATION = NUPHA,B226,187;%%
%``Continuum Limit And Improved Action In Lattice Theories. 2. O(N) Nonlinear Sigma Model In Perturbation Theory,''
Nucl.\ Phys.\ B {\bf 226},   205 (1983).
%%CITATION = NUPHA,B226,205;%%



\bibitem{Lepage:1992xa}
G.~Lepage and P.~Mackenzie,
%``On the viability of lattice perturbation theory,''
Phys.\ Rev.\ D {\bf 48}, 2250 (1993).
%[arXiv:hep-lat/9209022].
%%CITATION = HEP-LAT 9209022;%%


\bibitem{Griffiths:1983ah}
L.~A.~Griffiths, C.~Michael and P.~E.~Rakow,
%``Mesons With Excited Glue,''
Phys.\ Lett.\ B {\bf 129}, 351 (1983).
%%CITATION = PHLTA,B129,351;%%


\bibitem{Perantonis:1990dy}
S.~Perantonis and C.~Michael,
%``Static Potentials And Hybrid Mesons From Pure SU(3) Lattice Gauge Theory,''
Nucl.\ Phys.\ B {\bf 347},   854 (1990).
%%CITATION = NUPHA,B347,854;%%


\bibitem{Bernard:1997ib}
C.~Bernard {\it et al.}  [MILC Collaboration],
%``Exotic mesons in quenched lattice QCD,''
Phys.\ Rev.\ D {\bf 56}, 7039 (1997).
%[arXiv:hep-lat/9707008].
%%CITATION = HEP-LAT 9707008;%%



\bibitem{Lacock:1996ny}
P.~Lacock, C.~Michael, P.~Boyle and P.~Rowland  [UKQCD
Collaboration],
%``Hybrid mesons from quenched QCD,''
Phys.\ Lett.\ B {\bf 401}, 308 (1997).
%[arXiv:hep-lat/9611011].
%%CITATION = HEP-LAT 9611011;%%


\bibitem{McNeile:1998cp}
C. Bernard {\it et al.}, [MILC Collaboration],
%``Exotic meson spectroscopy from the clover action at beta = 5.85 and  6.15,''
Nucl.\ Phys.\  B(Proc. Suppl.){\bf 73},  264 (1999).
%[arXiv:hep-lat/9809087].
%%CITATION = HEP-LAT 9809087;%%



\bibitem{Manke:1998qc}
T.~Manke {\it et al.}  [CP-PACS Collaboration],
%``Hybrid quarkonia on asymmetric lattices,''
Phys.\ Rev.\ Lett.\  {\bf 82}, 4396 (1999).
%[arXiv:hep-lat/9812017].
%%CITATION = HEP-LAT 9812017;%%


\bibitem{Juge:1999ie}
K.~Juge, J.~Kuti and C.~Morningstar,
%``Ab initio study of hybrid anti-b g b mesons,''
Phys.\ Rev.\ Lett.\  {\bf 82},  4400 (1999).
%[arXiv:hep-ph/9902336].
%%CITATION = HEP-PH 9902336;%%


\bibitem{Drummond:1999db}
I.~Drummond {\it et al.},
%N.~A.~Goodman, R.~R.~Horgan, H.~P.~Shanahan and L.~C.~Storoni,
%``Spin effects in heavy hybrid mesons on an anisotropic lattice,''
Phys.\ Lett.\ B {\bf 478},  151 (2000).
%[arXiv:hep-lat/9912041].
%%CITATION = HEP-LAT 9912041;%%




\bibitem{Mei:2002ip}
Z.~H.~Mei and ~X.~Q.~Luo,
%`` EXOTIC MESONS FROM QUANTUM CHROMODYNAMICS WITH IMPROVED GLUON AND QUARK ACTIONS ON THE ANISOTROPIC LATTICE,''
Int.\ J.\ Mod.\ Phys.\ A {\bf 18}, 5713 (2003).
%[arXiv: hep-lat/0206012  ].
%%CITATION = HEP-LAT 0206012;%%


\bibitem{Bernard:2003jd}
  C.~Bernard {\it et al.}, [MILC Collaboration],
  %``Lattice calculation of 1-+ hybrid mesons with improved Kogut-Susskind
  %fermions,''
  Phys.\ Rev.\ D {\bf 68}, 074505 (2003).
%  [arXiv:hep-lat/0301024].
  %%CITATION = HEP-LAT 0301024;%%


\bibitem{Hedditch:2005zf}
  J.~N.~Hedditch {\it et al.},
  %W.~Kamleh, B.~G.~Lasscock, D.~B.~Leinweber, A.~G.~Williams and J.~M.~Zanotti,
  %``1-+ exotic meson at light quark masses,''
 Phys.\ Rev.\ D {\bf 72}, 114507 (2005).
% [arXiv:hep-lat/0509106].
  %%CITATION = HEP-LAT 0509106;%%


\bibitem{Michael:2003xg}
  C.~Michael,
  %``Hybrid mesons from the lattice,''
  hep-ph/0308293, and refs. theirin.
  %%CITATION = HEP-PH 0308293;%%



\bibitem{Thacker:1991ej}
B.~Thacker and G.~Lepage,
%``HEAVY QUARK BOUND STATES IN LATTICE QCD,''
Phys.\ Rev.\ D {\bf 43}, 196 (1991).
%%CITATION = PR,D43,196;%%


\bibitem{Lepage:1992ej}
G.~P.~Lepage {\it et al.},
%`` IMPROVED NONRELATIVISTIC QCD FOR HEAVY QUARK PHYSICS,''
Phys.\ Rev.\ D {\bf 46},  4052 (1992).
%[arXiv:hep-lat/9205007 ].
%%CITATION = HEP-LAT 9205007;%%


\bibitem{Trottier:1997ej}
H.~D.~Trottier,
%`` QUARKONIUM SPIN STRUCTURE IN LATTICE NRQCD,''
Phys.\ Rev.\ D {\bf 55}, 6844 (1997);
%[arXiv:hep-lat/9611026  ].
%%CITATION = HEP-LAT 9611026;%%
%\bibitem{Shakespeare:1997ej}
N.~H.~Shakespeare and H.~D.~Trottier,
%`` TADPOLE - IMPROVED SU(2) LATTICE GAUGE THEORY;TADPOLE RENORMALIZATION AND RELATIVISTIC CORRECTIONS IN LATTICE NRQCD,''
Phys.\ Rev.\ D {\bf 58}, 034502 (1998); $\emph{ibid.}$ {\bf 59},
014502 (1999).
%[arXiv:hep-lat/9803024; hep-lat/9802038 ].
%%CITATION = HEP-LAT 9803024; 9802038;%%


\bibitem{El-Khadra:1997ej}
A.~X.~El-Khadra, ~A.~S.~Kronfeld and P.~B.~Mackenzie,
%`` MASSIVE FERMIONS IN LATTICE GAUGE THEORY,''
Phys.\ Rev.\ D {\bf 55}, 3933 (1997).
%[arXiv: hep-lat/9604004 ].
%%CITATION = HEP-LAT 9604004;%%



\bibitem{Sheikholeslami:1985ij}
B.~Sheikholeslami and R.~Wohlert,
%``Improved Continuum Limit Lattice Action For QCD With Wilson Fermions,''
Nucl.\ Phys.\ B {\bf 259}, 572 (1985).
%%CITATION = NUPHA,B259,572;%%



\bibitem{Klassen:1998fh}
T.~R.~Klassen,
%``Non-perturbative improvement of the anisotropic Wilson QCD action,''
Nucl.\ Phys.\ Proc.\ Suppl.\  {\bf 73},   918 (1999).
%[arXiv:hep-lat/9809174].
%%CITATION = HEP-LAT 9809174;%%



\bibitem{Okamoto:2001jb}
M.~Okamoto {\it et al.}  [CP-PACS Collaboration],
%``Charmonium spectrum from quenched anisotropic lattice QCD,''
Phys.\ Rev.\ D {\bf 65},   094508 (2002).
%[arXiv:hep-lat/0112020].
%%CITATION = HEP-LAT 0112020;%%

\bibitem{Chen:2000ej}
P.~Chen,
%``Heavy quarks on anisotropic lattices: The charmonium spectrum,''
Phys.\ Rev.\ D {\bf 64},  034509 (2001).
%[arXiv:hep-lat/0006019].
%%CITATION = HEP-LAT 0006019;%%


\bibitem{Morningstar:1997prd}
C.~Morningstar and M.~Peardon,
%``Efficient glueball simulations on anisotropic lattices,''
Phys.\ Rev.\ D {\bf 56},  4043 (1997).
%[arXiv:hep-lat/9704011 ].
%%CITATION = HEP-LAT 97040114;%%
%\bibitem{Morningstar:1999prd}
%C.~Morningstar and M.~Peardon,
%``The glueball spectrum from an anisotropic lattice study,''
Phys.\ Rev.\ D {\bf 60},  034509 (1999).
%[arXiv:hep-lat/9901004].
%%CITATION = HEP-LAT 9901004;%%

\bibitem{Blum:2002ii}
T.~Blum,
  %``Lattice calculation of the lowest order hadronic contribution to the muon
  %anomalous magnetic moment. ((U)),''
  Phys.\ Rev.\ Lett.\  {\bf 91}, 052001 (2003).
%  [arXiv:hep-lat/0212018].
 %%CITATION = HEP-LAT 0212018;%%


\bibitem{Takahashi:2002it}
  T.~T.~Takahashi and H.~Suganuma,
  %``The gluonic excitation of the three-quark system in SU(3) lattice QCD,''
  Phys.\ Rev.\ Lett.\  {\bf 90}, 182001 (2003).
%  [arXiv:hep-lat/0210024].
  %%CITATION = HEP-LAT 0210024;%%

\bibitem{Gockeler:2003jf}
  M.~Gockeler {\it et al.}
  [QCDSF Collaboration],
  %``Generalized parton distributions from lattice QCD,''
  Phys.\ Rev.\ Lett.\  {\bf 92}, 042002 (2004).
%  [arXiv:hep-ph/0304249].
  %%CITATION = HEP-PH 0304249;%%

\bibitem{Lacock:1998be}
  P.~Lacock and K.~Schilling  [TXL collaboration],
  %``Hybrid and orbitally excited mesons in full QCD,''
  Nucl.\ Phys.\ Proc.\ Suppl.\  {\bf 73}, 261 (1999).
%  [arXiv:hep-lat/9809022].
  %%CITATION = HEP-LAT 9809022;%%


\bibitem{Manke:2001ft}
  T.~Manke {\it et al.}  [CP-PACS Collaboration],
  %``Hybrid quarkonia with dynamical sea quarks,''
  Phys.\ Rev.\ D {\bf 64}, 097505 (2001).
%  [arXiv:hep-lat/0103015].
  %%CITATION = HEP-LAT 0103015;%%


\bibitem{Milc} http://physics.utah.edu/$\sim$detar/milc/



\end{thebibliography}
\end{document}